# Effective temperatures out of equilibrium[*]

Leticia F. Cugliandolo

*Laboratoire de Physique Théorique de l'École Normale Supérieure*
*24 rue Lhomond, 75231 Paris Cedex 05, France and*
*Laboratoire de Physique Théorique et Hautes Energies, Jussieu*
*5 ème étage, Tour 24, 4 Place Jussieu, 75005 Paris France*

(January 24, 2017)

## Abstract

We describe some interesting effects observed during the evolution of nonequilibrium systems, using domain growth and glassy systems as examples. We breafly discuss the analytical tools that have been recently used to study the dynamics of these systems. We mainly concentrate on one of the results obtained from this study, the violation of the fluctuation-dissipation theorem and we discuss, in particular, its relation to the definition and measurement of effective temperatures out of equilibrium.

One of the major challenges in physics is to understand the behaviour of systems that are far from equilibrium. These systems are ubiquitous in nature. Some examples are phase separation, systems undergoing domain growth, all types of glasses, turbulent flows, systems driven by non-potential forces, etc. All these systems are "large" in the sense that they are composed of many, $N \to \infty$, dynamic degrees of freedom. Apart from succeeding in predicting the time evolution of their macroscopic properties, one would like to know which, if any, of the thermodynamic notions apply to these nonequilibrium cases.

Systems undergoing domain growth, or phase separation, provide the best known example of a nonequilibrium evolution.[1] Take for instance a magnetic system with ferromagnetic interactions in contact with a thermal bath. If the bath temperature is very high the sample is in its paramagnetic phase and the magnetic moments, or spins, point in random directions. If one next cools down the bath, and hence the sample, through a transition temperature $T_c$, the system enters the low temperature phase and starts forming *domains* or islands of the two ordered phases, say up and down. For definiteness, let us fix the final temperature to be $0 < T < T_c$. At any time $t_w$ after crossing the transition at the initial time, two types of dynamics appear: (i) *fast* fluctuations of some spins, due to thermal fluctuations,

---

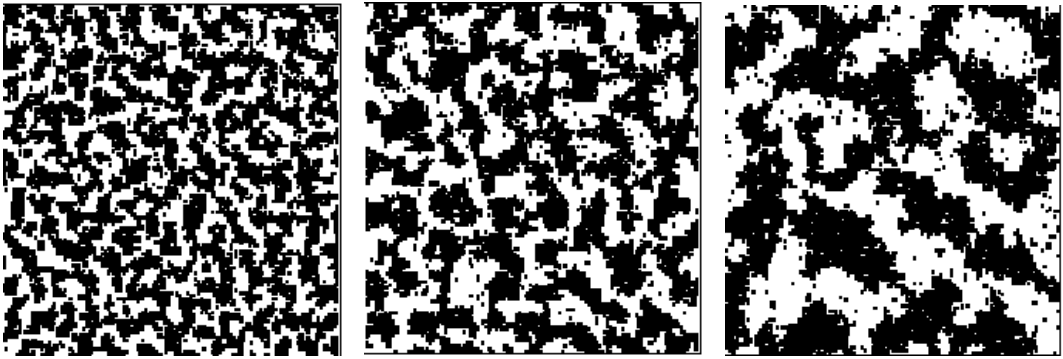

FIG. 1. Three two-dimensional slices of a domain-growth system at three different waiting times $t_o < t_{w1} < t_{w2} < t_{w3}$. See the text for a discussion.

inside the otherwise fully ordered domains; (ii) *slow* motion of the domain walls leading to the growth of the averaged domain size $L(t_w)$. If the size of the sample is infinite, in real life very large, the nonequilibrium domain-growth process can take so long that the sample simply does not equilibrate in the time-window that is accessible experimentally. In other words, below the critical temperature $T_c$ one always has $\tau_{\text{OBS}} < \tau_{\text{EQ}}$ with $\tau_{\text{OBS}}$ the observation time and $\tau_{\text{EQ}}$ the equilibration time. The two types of dynamics itemized above are clearly seen in Fig. 1 where three two-dimensional slices of a system undergoing domain-growth are displayed. The pictures are obtained at increasing waiting times after the quench. One sees the domains growing as well as the existence, in each of the snapshots, of some reversed spins inside the otherwise ordered domains.

Scaling arguments have been extensively used to describe the dynamics below $T_c$; they are based on the assumption (sometimes derivation) of the evolution of the averaged domain size $L(t_w)$ and on further proposals for the space and time-dependence of the correlation functions.

Understanding the physics of glassy materials is perhaps a problem of intermediate difficulty.[2,3] Glassy materials can be of very many different types; one has for instance structural glasses,[2] orientational glasses,[4] spin-glasses,[5,6] plastics,[7] gels and clays,[8] glycerol,[9] etc. Their hallmark is that below some transition range they also fall out of equilibrium.

The easiest way of preparing a glassy system is again through an annealing.[2] This is implemented by decreasing the temperature of the bath with a given cooling-rate. Take for example the case of a molecular liquid. For high enough bath temperatures the sample is in its liquid phase and it achieves equilibrium with the bath. At an intermediate bath-temperature range the liquid avoids the crystallization transition, enters a metastable phase and becomes a *super-cooled liquid*, that is to say a liquid with some peculiar properties as, for example, a extremely high viscosity. At an even lower bath-temperature the liquid cannot follow the pace of the annealing and falls out of equilibrium, it becomes a *glass*. If one stops the annealing at any temperature below this range the system stays in its glassy phase for practical purposes forever and is typically an amorphous solid. We talk about a "transition range" since the transition might not be clearcut but depend on the cooling-rate. Actually one can form a glass of probably any substance by choosing a fast enough cooling-rate. Many other routes to the glassy phase are also possible.

There have been proposals to describe the evolution of some glasses, notably spin-glasses, with scaling arguments based on domain growth ideas.[10,11] The assumption is that the glassy dynamics is simply given by the growth of domains of two competing ground states. However, it has been very difficult to prove (or disprove!) either experimentally or numerically that this is indeed the scenario: no "ordered structures" have been identified in general as the growing phases.

Thus, domain growth, phase separation and glassy materials are all "self-sustained"[1] out of equilibrium systems. If one follows their time-evolution, keeping all parameters fixed, in particular the bath-temperature, some of the main features observed during their nonequilibrium evolution are:

*Slow dynamics.* The evolution is very slow. "One-time quantities", as the energy-density, approach their asymptotic limit with some slow decaying function, say power law, logarithmic or more complicated. It is very important to notice though that even if these one-time quantities can get very close to their asymptotic values, this does not mean that the systems *get frozen* in a metastable state: they are not equilibrated in a restricted region of phase space characterised by these asympotic values. This is most clearly demonstrated by the measurement of "two-time quantities".

*Two-time quantities and physical aging.* The measurement of these quantities prove that, even if one-time quantities approach a limit, the system is still changing in an important way.

One can distinguish two types of two-time quantities. Those measured during the free evolution of the system, quantifying the spontaneous fluctuations, such as any two-time correlation function, and those measured after applying a small perturbation, such as any response function. These quantities depend on both times involved in the measurement and not only on the time-difference. This shows that the systems neither are equilibrated with the bath nor have approached equilibrium in any metastable state. They are indeed rather far from equilibrium.

The measurement of the spontaneous fluctuations is quite easy to implement in a numerical simulation.[12] One prepares the sample at an initial time $t_o$ and lets it evolve until a waiting time $t_w$ when the system configuration is recorded. One then lets the sample further evolve and computes, at all subsequent times $t \equiv \tau + t_w$, the correlation function between the reference configuration at $t_w$ and the configurations at $\tau + t_w$. These curves depend on both $t_w$ and $\tau$ and they are not invariant under time translations showing that the system is out of equilibrium. Furthermore, the decay as a function of $\tau$ is slower the longer $t_w$. This is the phenomenon called *physical aging*: the younger (older) the sample the faster (slower) the decay.

The result of the measurement of a local auto-correlation function is very easy to visualize for a domain growth. Take again the case of ferromagnetic domain growth. the dynamic variables are the Ising spins that we encode in a time-dependent $N$-dimensional vector $\boldsymbol{\phi}(t) = (\phi_1(t), \ldots, \phi_N(t))$ (the index $i = 1, \ldots, N$ labels the spins) and the local auto-correlation function is just the scalar product of two configurations evaluated a different times, $NC(t, t_w) = \langle \boldsymbol{\phi}(t) \cdot \boldsymbol{\phi}(t_w) \rangle$. For Ising spins, the auto-correlation function is normalized to one at equal times. A departure from one measures how different are any two configurations as those shown in Fig. 1. For any fixed $t_w$ the auto-correlation has two distinct regimes depending on the time-difference $\tau \equiv t - t_w$. Let us choose a waiting time $t_{w1}$ and plot $C(\tau + t_{w1}, t_{w1})$ as a function of $\tau$. The curve has a first fast decay from one to a bath-temperature dependent value $q_{EA}(T)$ (the Edwards-Anderson temperature-dependent parameter). This corresponds to the decorrelation associated to the fast flipping of the spins inside the domains. In this regime $\tau$ is small compared to an increasing function of the domain size $g(L(t_w))$. When $\tau$ increases and becomes of the order of $g(L(t_w))$ one starts seeing the motion of the domain-walls, i.e. the growth of the domains, and the decay slows down. If one repeats this calculation choosing a longer waiting-time $t_{w2} > t_{w1}$, and its associated reference configuration, one observes that the first decay is identical to the one for $t_{w1}$ though it lasts for longer, and that the second regime is notably slower that the one associated to $t_{w1}$. These features can be easily understood. While $\tau$ is smaller than $g(L(t_w))$ the dynamics takes place only inside the domains as thermal fluctuations. The domain walls are ignored and the correlation behaves as if the system were a patchwork of the two equilibrium states. The correlation function decay is then independent of the waiting-time and approaches $q_{EA}(T) = m_{\text{EQ}}^2(T)$. However, after a time-difference of the order of $g(L(t_w))$ the system realizes it has domain walls, the subsequent decay is associated to the motion of the walls and is nonequilibrium in nature. The decay gets slower the longer the $t_w$ simply because the size of the domains reached at $t_w$ is larger.

---

[1]In the sense that no external perturbation is keeping them far from equilibrium.

For glassy systems the correlation functions have exactly the same qualitative behaviour though, as already mentioned, it is not easy to decide if there is any type of order growing. Plots like the one displayed in Fig. 2 have been obtained for an impressive number of glassy models of different nature. Some of them are the 3D Edwards-Anderson model,[12] a polymer melt,[13] a polymer in a random potential,[14] a binary Lennard-Jones mixture,[15] etc. Furthemore, the kind of curves were found in several sandpile models and other kind of systems.[16]

The measurement of "dc-response" functions or, more precisely, integrated dc-responses is what is usually done experimentally. The starting procedure is similar to the precedent: one prepares the sample at an initial time and lets it freely evolve until $t_w$. At this waiting time one applies a small, constant, perturbation and then measures the associated integrated response of the system as a function of $\tau$ for different $t_w$. For example, experimenting with spin-glasses one applies a small magnetic field and measures the increase of magnetization,[6] manipulating with polymer glasses one applies a stress and measure the tensile creep compliance,[7] etc. In all cases, the integrated responses are studied as functions of $t_w$ and $\tau$ and they all show aging effects that manifest in a similar way as in the correlation measurement. There is a first increase of the time-integrated susceptibility towards a value $\chi_1(T)$ that does not depend on the waiting-time while there is a second increase of the time-integrated susceptibility towards the equilibrium value $\chi_{\text{EQ}}(T)$ that is waiting-time dependent.

This can again be simply visualize in the domain growth problem. The first regime corresponds to the response of the spins inside the domains, i.e. to the response of the full system taken to be roughly a patchwork of independent equilibrium states. The second regime instead corresponds to the response of the domain walls. Since their density decreases as time elapses, one expects this nonequilibrium response to vanish in the long witing-time limit.

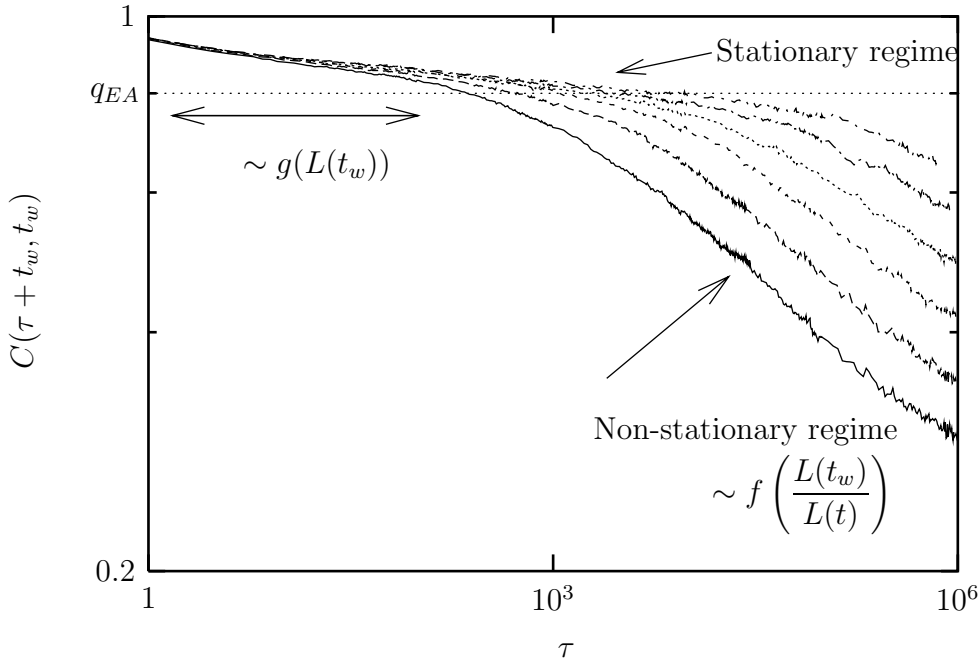

FIG. 2. The local auto-correlation function for wait-times $t_{w1} < t_{w2} < t_{w3} < t_{w4} < t_{w5} < t_{w6}$. See the text for a discussion of the different regimes.

"Ac-response" measurements are also usually performed experimentally.[4,9] In these experiments one applies an ac small field of fixed frequency $\omega$ at the initial time $t_0$ and keeps it applied until the measuring time $t_w$. The relation with the previous results is given by identifying $\omega \sim 1/\tau$. The out of equilibrium character of the evolution is given by an explicit $t_w$ dependence in the relaxation of the in and out of phase susceptibilities.

It is important to notice that these effects are *physical aging* as opposed to *chemical aging*. Physical aging is totally reversible: it suffices to heat the sample above the transition range and cool it back again below it to recover a fully rejuvenated system.

The comparison of spontaneous fluctuation, e.g. a correlation function, to induced fluctuations, measured by its associated response function, is well established for systems evolving in equilibrium. Indeed, this relation involves the temperature of the bath and it is called the *fluctuation-dissipation theorem*. However, for systems that are far from equilibrium this relation does not necessarily hold.

The problem of reaching a theoretical understanding of nonequilibrium physics is important both from a practical and a theoretical point of view. It is obvious that for some applications, one would like to predict the time-evolution of the samples with great precision and avoid undesired changes that depend on the, sometimes unknown, age of the samples. From an analytic point of view, domain-growth and glassy materials are a pattern of out of equilibrium systems whose properties one could try to capture with simple models or simplified approaches to more complex models. The predictions thus obtained can then be experimentally (or numerically) tested in real systems. Importantly enough, one could also try to extend some of these predictions to other nonequilibrium systems such as those externally driven. Some connections between glassy systems and driven systems are discussed in Ref. [17].

How can one modelise domain growth or glassy systems? The "microscopic" constituents and interactions and, consequently, the microscopic models differ from glass to glass. Spin-glasses are composed of magnetic impurities (spins) that occupy fixed random positions in the sample and interact via RKKY interactions; polymer glasses are composed of polymers (strings) that interact via potentials of (oversimplifying) Lennard-Jones type. In the former case, there is quenched (time-independent) disorder in the system that is associated to the random positions of the spins, that give rise to random interactions between them (the RKKY interactions oscillate very rapidly with the distance between the magnetic impurities and change sign in an almost random manner). In the latter case nothing can be interpreted as being quenched disorder. Though one could expect that these two systems (and other type of glasses) behave very differently, the experimental results as well as the recent theoretical developments show that their dynamical behaviour is indeed rather similar. In other words, one can identify certain quantities that have the same qualitative behaviour.

In order to describe the dynamic evolution of a *classical* system in contact with an environment one starts by identifying the relevant variables of the system that evolve in time. One then proposes a Langevin equation, with noise and friction mimicking the coupling of the system to the thermal bath, to determine the time-evolution of the time-dependent variables. These variables are the (soft) spins in the case of a spin-glass, the monomer positions in the case of the polymer glass, etc. This procedure leads to a set of $N$, the number of dynamic variables in the system, coupled differential equations of second (if inertia is included) or first (if inertia is neglected) order. Obviously, this huge system cannot be solved and one has to resort to some alternative method to further advance in the analysis.

Indeed, one would like to obtain information about "macroscopic" quantities, that can be related to experimental measurements, instead of following the erratic motion of any microcopic variable. The two-point correlation or response functions are macroscopic quantities we would like to monitor. A well-known theoretical method, known under the name of Martin-Siggia-Rose (MSR) formalism,[18] allows us to obtain a generating functional, and from it the Schwinger-Dyson integro-differential equations, for these quantities. This method can be applied in complete generality, to any classical model. For models with non-linear interactions of finite range it reccuires the calculation of an infinite series of diagrams that one cannot in general resum and express in terms of two-point functions in an explicit form.[19] One then faces the problem of choosing some approximation scheme to simplify this series expansion.

Before discussing how to deal with this problem, let us describe the formalism used to study a *quantum* system in contact with an environment. The Schwinger-Keldysh closed-time path (CTP) formalism was developed to monitor the nonequilibrium time-evolution of a quantum system, and to obtain information about two-time quantities.[20] The environment is usually modelized by a set of harmonic oscillators (infinitely many for each variable in the system) with a spectral distribution of frequencies.[21] The coupling of system and bath is usually chosen to be linear but of course more general situations can be considered. In this way, one obtains the CTP generating functional that, as in the classical case, involves a series expansion that, in general, cannot be obtained explicitly. (The classical limit of the CTP generating functional is the MSR generating functional.)

Typically, two routes are followed to approximate these generating functionals. They are equally applicable in the classical and quantum case and are the following:[3]

- The microscopic models, namely the starting Hamiltonians, are simplified in such a way that the construction can be carried through and that explicit equations can be derived. This is the choice made when one uses, for example, the large $N$ limit of a $O(N)$-model to describe domain growth, fully connected spin models to describe spin-glasses or when one embeds a finite dimensional manifold in an infinite dimensional space to describe an interface motion, the motion of a polymer in a random medium.

- The microscopic models are realistic but some approximation scheme is chosen to select, from the infinite series, a still infinite subset of diagrams that can be resummed to yield an explicit set of dynamic equations for correlations and responses. Many such recipes exist in the literature,[22–24] some of them are the mode-coupling approximation, the direct interaction approximation, the self-consistent screening approximation, etc.

These two procedures yield the same "form" of coupled integro-differential causal equations. Actually, in some cases one can show that a simplified microscopic model with infinite range interactions (e.g. the $p$ spin-glass model) yields the same dynamic equations that an approximation scheme (e.g. the mode coupling appoximation) applied to a more realistic model for a glassy material.[25,26] The structure of these equations is always the same: there might be a second time-derivative term if there is inertia, some terms describe the interaction of the system with the bath and some other integral terms describe the interactions in the system (through the self-energy and vertex). It is the explicit form of the self-energy and vertex that is selected by the model or the approximation.

Once one has the equations governing the evolution of the two-time quantities, for all values of the parameters, the question then arises as to which is the phenomenology that they describe.

A combination of analytic and numeric methods are used to study these equations. One can attempt a numerical solution taking advantage of the fact that they are causal. The solution shows that they present a dynamic phase transition at a temperature $T_d$. Above $T_d$, the solution reaches, in the long waiting-time limit, a stationary form. All two-time correlations and responses are functions of the time-difference only and are related through the fluctuation-dissipation theorem. The high-temperature dynamic equations for spin-glass were studied in detail by Sompolinsky and Zippelius[27] for spin-glass models, by Götze and collaborators[28] for glass-models and the relation between these two was signalled and investigated by Kirkpatrick, Thirumalai and Wolynes[25] in a series of beautiful papers.

Below $T_d$, a drawback of the numerical method is that, due to the slowness of the dynamics and the memory of the system, one cannot reach very long time intervals. The numerical solution gives us hints about the structure of the solution but does not give us extremely precise information about more detailed features such as the two-time scaling laws, etc. Nevertheless, the numerical solution sufficed to show that below $T_d$ two-time functions start depending on the waiting-time and that aging is captured by these equations.[29]

Below $T_d$, and in the asymptotic limit of long waiting-time, an analytical solution was developed first for the $p$ spin-glass model[29] and later for other mean-field disordered models such as Sherrington-Kirkpatrick[30] or the motion of manifolds in infinite dimensional random potentials.[31,32]

One of the main ingredients of this solution[29,30] concerns the *fluctuation-dissipation theorem* that relates, in equilibrium, the spontaneous to the induced fluctuations. Indeed, if one follows the dynamics of a classical system that is in equilibrium with a bath, one can easily show that

$$R(t,t') \equiv \left.\frac{\delta\langle O(t)\rangle}{\delta h(t')}\right|_{h=0} = \frac{1}{T}\frac{\partial}{\partial t'}\langle O(t)O(t')\rangle \ \theta(t-t') = \frac{1}{T}\frac{\partial}{\partial t'}C(t,t') \ \theta(t-t') , \qquad (1)$$

with $O(t)$ any observable taken to have zero mean for simplicity and $h$ an infinitesimal field acting a time $t'$ that modifies the energy of the system according to $V \to V - hO$ and that is not correlated with the equilibrium configuration of the system.

In the glassy phase, this relation does not hold. This does not come as a surprise since the equilibrium condition under which it can be proven does not apply. What really comes as a surprise is that the modification of the relation between response and correlation takes a rather simple form for domain-growth and glassy systems.

A way to quantify the modification of FDT in the out of equilibrium phase and to use it to classify different systems is the following. Let us integrate the response function over a time-interval going from a waiting-time $t_w$ to a final time $t$:

$$\chi(t,t_w) = \int_{t_w}^{t} dt' R(t,t') . \qquad (2)$$

This yields a time-integrated susceptibility that is exactly what is measured experimentally. Next, we compare this integrated-susceptibility to the auto-correlation function. *In equilibrium*, one can use FDT to show that

$$\chi(t,t_w) = \frac{1}{T}\left(C(t,t) - C(t,t_w)\right) . \qquad (3)$$

Hence, if one draws a plot of $\chi$ against $C$, for increasing $t_w$, using $\tau = t - t_w$ as a parameter, in the large $t_w$ limit the plot will approach a straight line of slope $-1/T$ joining $(\lim_{t\to\infty} C(t,t), 0)$ and $(0, \chi_{\text{EQ}})$. From now on and without loss of generality we take $\lim_{t\to\infty} C(t,t) = 1$. Any departure from this straight line signals a modification of FDT and a departure from equilibrium.

The analytic solution of simplifed models shows that, in the *nonequilibrium* phase, this construction converges to a limiting curve given by

$$\lim_{t_w\to\infty, C(t,t_w)=C} \chi(t,t_w) = \frac{1}{T_{\text{EFF}}(C)}(1-C) \qquad (4)$$

where $T_{\text{EFF}}(C)$ is a function of the correlation $C$. We shall discuss the notation and justify the name of this function below. In the large $t_w$ limit two distinct regimes develop in the $\chi$ vs $C$ curve. There is a first straight line of slope $-1/T$, joining $(1,0)$ and $(q_{EA}(T), \chi_1(T))$. This characterises what is called the FDT regime. The straight line then breaks and the $\chi$ vs $C$ curve goes on in a different manner. The subsequent behaviour depends on the model. Indeed, three families have been identified:

- Models describing domain growth like, for example, the $O(N)$ model in $D$ dimensions in the large $N$ limit. In this case, one follows the local correlation $NC(t,t_w) \equiv \langle \boldsymbol{\phi}(\boldsymbol{x},t) \cdot \boldsymbol{\phi}(\boldsymbol{x},t_w)\rangle$ and its associated local susceptibility. The plot for $C \leq q_{EA}(T)$ is flat.[34] The susceptibility gets stuck at its value $\chi_1(T)$ while the correlation continues decreasing towards zero. The same result holds for the Ohta-Jasnow-Kawasaki approximation to the $\lambda\phi^4$ model of phase separation.[35]

- Models describing structural glasses like, for example, the so-called $F_{p-1}$ models of the mode-coupling approach or the $p$ spin-glass models. In this case the $\chi$ vs $C$ plot, for $C \leq q_{EA}(T)$, is a straight line of slope larger than $-1/T$.[29]

- Models describing spin-glasses like, for example, the Sherrington-Kirkpatrick model. In this case the $\chi$ vs $C$ plot, for $C \leq q_{EA}(T)$, is a non-trivial curve.[30]

This "classification" in three families has been checked numerically for more realistic models. Many numerical simulations using either Montecarlo (MC) techniques or molecular dynamics (MD) have shown that several models fall into the expected cathegories. Some models belonging to the first group are the $2D$ Ising model with conserved and non-conserved order parameter,[36] the site diluted ferromagnet and the random field Ising model[37] and the $2D$ Ising model with ferromagnetic exchange and antiferromagnetic dipolar interactions.[38] The binary Lennard-Jones mixtures are a standard model for the glass transtion. Both MC and MD simulations show that they belong to the second class.[39] Finally, MC simulations of finite dimensional spin-glass models, the three and four dimensional Edwards-Anderson model, yield the third kind of behaviour.[40] Another particularly interesting problem, relevant for the physics of dirty superconductors,[32] is the one of a manifold diffusing in a random potential. The analytic prediction using an infinite dimensional embedding space depends on the nature of the quenched random potential, namely on it being short or long range correlated.[33] This prediction is partially confirmed by the simulations in finite dimensional transverse space with the proviso of a very interesting modification that is not captured by the infinite dimensional approach.[41] Besides, numerical simulations of lattice-gas modes with kinetic constraints[42] and sandpile models[43] also show FDT violations.

Once this modification of FDT in the nonequilibrium situation is identified, several questions arise, all connected with the initial purpose of checking which thermodynamic concepts can be applied, perhaps after some modifications, to the nonequilibrium case. In the following we discuss three interesting issues.

- Why is there always a two-time regime, when $C$ first decays from its equal times value to $q_{EA}(T)$, where FDT holds?

For the domain-growth problem the presence of this piece is easy to justify. In this time-scale one only sees the dynamics and the effect of the perturbation inside the domains. Since one can then ignore the presence of domain walls, the equilibrium relation between correlation and response is expected to hold. Of course, one cannot easily extend this argument to a more general situation. There is however a totally general reason for having FDT when $C \geq q_{EA}(T)$ and it is the following.

For any system in contact with an environment, with bounded correlation functions and without non-potential forces[2] the departure from FDT is bounded by[44]

$$|T\chi(t,t_w) - C(t,t) + C(t,t_w)| \leq K \int_{t_w}^{t} dt' \left(-\frac{1}{\gamma N}\frac{d\mathcal{H}(t')}{dt'}\right)^{1/2} \quad (5)$$

where $K$ is a finite constant and $\gamma$ the friction coefficient that characterises the coupling to the bath. The Kubo $\mathcal{H}$-function is defined as[46]

$$\mathcal{H}(t') \equiv \int d\boldsymbol{\phi} d\dot{\boldsymbol{\phi}} \, P(\boldsymbol{\phi},\dot{\boldsymbol{\phi}},t) \left(T \ln P(\boldsymbol{\phi},\dot{\boldsymbol{\phi}},t) + V(\boldsymbol{\phi}) + \frac{m\dot{\boldsymbol{\phi}}^2}{2}\right) , \quad (6)$$

with $P(\boldsymbol{\phi},\dot{\boldsymbol{\phi}},t)$ the time-dependent probability distribution, $V(\boldsymbol{\phi})$ the potential energy and $m$ a mass. The $\mathcal{H}$-function satifies $\dot{\mathcal{H}} \leq 0$ for all times and it vanishes only for the canonical distribution.

From this bound one sees that if $\mathcal{H}(t)$ falls to zero faster than $1/t$ no FDT violations are allowed in the long $t_w$ limit since the right-hand-side in Eq. (5) vanishes. Instead, if $\mathcal{H}(t)$ falls to zero in a slower manner, FDT is imposed by the bound for small time-differences but violations are allowed for longer time-differences. This argument proves that there is always a region of correlations close to $C = 1$ in which FDT holds, even for a system that is not close to equilibrium.

- Can one identify the slope of the plot with an inverse effective temperature and call it $-1/T_{\text{EFF}}(t,t_w) = -1/T_{\text{EFF}}(C)$?

About ten years ago, in the context of weak-turbulence, Hohenberg and Shraiman[47] proposed to define an effective temperature through the departure from FDT. However, a detailed analysis of this quantity and its properties was not given in this reference.

Indeed, one expects that any quantity to be defined as a nonequilibrium effective temperature must fulfill the requirements associated to the intuitive idea of temperature. The first property to check is if this effective temperature is measurable by a thermometer that is weakly coupled to the system, in a statistical manner, at any chosen waiting time.[48] This property can be proven by studying the time-evolution of the thermometer coupled to $M$ identical copies of the system, all of age $t_w$, and by verifying that this equation becomes a Langevin equation in the presence of a thermal bath characterised by a *coloured noise* with correlation given by the system's correlation and response given by the system's response.

---

[2]Other bounds can be found if diffusion and/or non-potential forces are allowed.[44,45]

Thus, if the system has several time-scales characterized by different values of the effective temperatures[3]

$$C(t, t_w) = C^{\text{FDT}}(t, t_w) + C^{(1)}(t, t_w) + C^{(2)}(t, t_w) + \ldots \quad (7)$$
$$R(t, t_w) = R^{\text{FDT}}(t, t_w) + R^{(1)}(t, t_w) + R^{(2)}(t, t_w) + \ldots \quad (8)$$

with

$$R^{\text{FDT}}(t, t_w) = \frac{1}{T}\frac{\partial}{\partial t_w}C^{\text{FDT}}(t, t_w)\,\theta(t - t_w) \qquad R^{(i)}(t, t_w) = \frac{1}{T^{(i)}}\frac{\partial}{\partial t_w}C^{(i)}(t, t_w)\,\theta(t - t_w) \quad (9)$$

one can select which value $T^{(i)}$ is measured by choosing the internal time-scale of the thermometer. Say, for example, that the thermometer is a harmonic oscillator of internal frequency $\omega_o$. Then, one chooses the system time-scale to be explored, and hence the value of the effective temperature to be measured, by comparing $\omega_o$ to $t_w$.

Many desirable "thermodynamic" properties of $T_{\text{EFF}}$ defined in this way can also be checked,[48] for example:

(i) $T_{\text{EFF}}$ controls the direction of heat flows.

(ii) $T_{\text{EFF}}$ controls partial equilibrations between observables in a system that evolve in the same time-scales and interact strongly enough.

(iii) Let us take two different glasses, in contact with a single bath of temperature $T$. These glasses are constructed in such a way that when they are not in contact each of them has a piecewise $T_{\text{EFF}}(C)$ of the form

$$T_{\text{EFF}}^{\text{SYST}\,1}(C) = \begin{cases} T & \text{if } C > q_{EA}^{(1)} \\ T^{(1)} & \text{if } C < q_{EA}^{(1)} \end{cases} \qquad T_{\text{EFF}}^{\text{SYST}\,2}(C) = \begin{cases} T & \text{if } C > q_{EA}^{(2)} \\ T^{(2)} & \text{if } C < q_{EA}^{(2)} \end{cases} \quad (10)$$

with $T^{(1)} \neq T^{(2)}$. One can then reproduce the experiment of setting two observables in contact by coupling these two systems through a small linear coupling between their microscopic variables. The result is that above a critical (though small) value of the coupling strength the two values the effective temperatures below $q_{EA}$ equal while below the same critical value of the coupling strength the values remain unaltered. One concludes that if the two observables interact strongly the systems arrange their time-scales in such a way to partially thermalise.

The presence of non trivial effective temperatures in glycerol out of equilibrium is presently being checked experimentally by Grigera and Israeloff.[49] Their results show that, at fixed measuring frequency $\omega_o \sim 8$ Hz, this system has an effective temperature $T_{\text{EFF}} > T = 180K$ until measuring times of at least $10^5$ sec, that is to say of the order of days! (Note that the bath temperature $T$ is below $T_c = 187$ K.)

---

[3]See Ref. [30] for a precise definition of two-time scales.

Further support to the notion of effective temperatures comes from the study of the effect of quantum fluctuations on the same family of models.[50] Below a critical line, that separates glassy from equilibrium phases, and in the slow dynamic regime, one finds violations of the quantum fluctuation dissipation theorem. These are characterised by the replacement of the bath temperature by an effective temperature $T_{\text{EFF}}(t, t_w)$. The effective temperature is again piecewise. It coincides with the bath-temperature $T$ when $C$ is larger than $q_{EA}$ and it is different when $C$ goes below $q_{EA}$. This nonequilibrium value has the nice property of being non-zero even at zero bath-temperature. Again, this result can be interpreted within the domain growth example. Whenever one looks at short time-differences with respect to the waiting-time one explores the quantum and thermal fluctuation in the bulk, i.e. one observes a quantum equilibrium dynamics that satisfies the quantum FDT. Instead, when $\tau$ is comparable to $g(L(t_w))$ one observes the domain wall motion. These are macroscopic objects for which quantum fluctuation do not have a strong effect. This can be seen, for example, in the form of the FDT violations: they look classical though with an effective temperature that depends on the strength of quantum fluctuations.

- Do effective temperatures in out of equilibrium systems emerge from a symmetry breaking?

In the classical case, one can study the structure of time-scales and effective temperatures with the help of the supersymmetric formulation of stochastic processes.[51,52] Indeed, it is well-known that the effective action in the MSR generating functional is invariant under a supersymmetric group (with a possible symmetry breaking due to the initial condition).

In the kind of glassy systems we deal with, there is a neat separation of time-scales in the long waiting-time limit. This allows us to separate the dynamics in the fast scale from the dynamics in the slow time-scales. The equation governing the slow time-scales, have an enlarged symmetry: they acquire an invariance under super-reparametrizations. The only solution that respects the large symmetry is a trivial, constant one. Hence, in order to have non-trivial dynamics in the long waiting-time limit, the system has to spontaneously break the super-reparametrization invariance. One can prove that the choice of effective temperatures is intimately related to the spontaneous breaking of this invariance.[53]

A similar analysis in the quantum case remains to be developed.

In conclusion, we have summarized some interesting features of the slow out of equilibrium dynamics of domain growth and glassy systems. We have explained why these features arise in the domain-growth case. A similar understanding has not been reached for glassy systems yet. With the purpose of developing a "visual" understanding of glassy physics, a careful analysis of the statistics and organisation of the configurations visited by a glass model during its nonequilibrium evolution is in order.

The use of simplified models or, alternatively, self-consistent approximations to more realistic ones have yielded a number of very interesting results and new predicitons. In particular, these models capture much of the aging phenomenology of glassy systems. Surprisingly enough, even puzzling effects of temperature cyclings during aging in spin-glasses, and the absence of these effects in other kind of glasses, can be described by fully-connected models.[54] Some of these new predictions, notably the modification of FDT, have been tested numerically and experiments are now being performed. Obviously, it is desirable to go beyond these approximations and study more realistic models in finite dimensions. This, however, is a very difficult task.

There have been innumerable attempts to define a temperature for an out of equilibrium system. In particular, in the context of glassy materials, a "fictive temperature" is often introduced to describe some of the experimental findings.[7] The effective temperature discussed in this article has the most welcome property of being measurable, hence being open to experimental tests. As far as we have checked the definition, it also has the welcome property of conforming to the common prejudices one has of a temperature. Of course there are still many open questions related to it. Just to mention one, let us say that it would be very interesting to extend the analytical experiment of "coupling a thermometer to a system" to the quantum case.

Acknowledgements
I wish to especially thank J. Kurchan with whom I have done much of the work on this subject and H. Castillo for suggestions concerning the preparation of this manuscript.